\documentclass[11pt]{article}

\usepackage{amsmath}
\usepackage{graphicx}
\usepackage{amsfonts}
\usepackage{amssymb}
\usepackage{epsfig}
\usepackage{color}
\usepackage{psfrag}
\usepackage{epstopdf}

\newcommand{\EQ}{\begin{equation}}
\newcommand{\EN}{\end{equation}}
\newcommand{\be}{\begin{equation}}
\newcommand{\ee}{\end{equation}}
\newcommand{\bea}{\begin{eqnarray}}
\newcommand{\eea}{\end{eqnarray}}

\newcommand{\nn}{\nonumber \\}

\begin{document} \setcounter{page}{0}
\topmargin 0pt
\oddsidemargin 5mm
\renewcommand{\thefootnote}{\arabic{footnote}}
\newpage
\setcounter{page}{0}
\topmargin 0pt
\oddsidemargin 5mm
\renewcommand{\thefootnote}{\arabic{footnote}}
\newpage
\begin{titlepage}
\begin{flushright}
\end{flushright}
\vspace{0.5cm}
\begin{center}
{\large {\bf On the phase diagram of the random bond $q$-state Potts model}}\\
\vspace{1.8cm}
{\large Gesualdo Delfino$^{1,2}$ and Noel Lamsen$^{1,2}$}\\
\vspace{0.5cm}
{\em $^1$SISSA -- Via Bonomea 265, 34136 Trieste, Italy}\\
{\em $^2$INFN sezione di Trieste, 34100 Trieste, Italy}\\
\end{center}
\vspace{1.2cm}

\renewcommand{\thefootnote}{\arabic{footnote}}
\setcounter{footnote}{0}

\begin{abstract}
\noindent
We consider the two-dimensional random bond $q$-state Potts model within the recently introduced exact framework of scale invariant scattering, exhibit the line of stable fixed points induced by disorder for arbitrarily large values of $q$, and examine the renormalization group pattern for $q>4$, when the transition of the pure model is first order. 
\end{abstract}
\end{titlepage}

\newpage
Gaining theoretical access to the critical properties of disordered systems with short range interactions has been a main challenge of statistical mechanics. Even in the two-dimensional case, for which infinite-dimensional conformal symmetry provided a huge amount of exact results for pure (i.e. not disordered) systems, analytic information remained confined to some perturbative limits. In this context, the two-dimensional random bond $q$-state Potts model attracted a special attention. In the first place, it was shown, rigorously in \cite{AW} and by renormalization arguments in \cite{HB}, that disorder softens, through the elimination of the discontinuity in the energy density, the first order transition that the pure model exhibits for $q>4$; a second order transition extending to infinite $q$ has then be expected. Since the random critical point of the model is perturbatively accessible only for $q\to 2$ \cite{Ludwig,DPP}, this expectation remained out of reach of analytic investigation. It was, however, supported by numerical studies, which at the same time pointed to a peculiar superuniversality (i.e. $q$-independence) of critical exponents along the random critical line \cite{CFL,DW,KSSD}, a suggestion no longer considered after that a clear $q$-dependence of the magnetization exponent $\beta$ was numerically exhibited in \cite{CJ}. 

Somehow surprisingly, it has been recently shown \cite{random} that critical lines of two-dimensional models with quenched disorder can be determined exactly within the scale invariant scattering method introduced in \cite{paraf}. Relying on symmetry, the approach is general and has been applied to the disordered $O(N)$ model \cite{DL1,DL2} and Potts model \cite{random,DT2}. For the latter, it shows that the line of stable renormalization group fixed points induced by disorder indeed extends beyond $q=4$ until $q=\infty$, but also unveils a subtle mechanism allowing for a superuniversal correlation length exponent $\nu$ and a $q$-dependent $\beta$. This finally accounts for the persisting -- and puzzling -- indications \cite{CFL,DW,KSSD,CJ,CB2,OY,JP,Jacobsen_multiscaling,AdAI} that $\nu$ does not show any appreciable deviation from the Ising value up to $q=\infty$. 

In this paper we look more closely into the problem of the softening of the transition for $q>4$. Indeed, having found the line of stable fixed points that extends to arbitrarily large values of $q$ is not enough from the point of view of the renormalization group. Within the space of parameters, this line will be the large distance limit of a second order transition surface that has to originate somewhere else. For $q\leq 4$ it originates from the line of fixed points of the pure model, which is unstable under the action of disorder. Where does the surface originate from for $q>4$? Since the correlation length is infinite on the second order surface, the latter cannot originate from the first order transition line of the pure model, along which the correlation length is finite. Given its generality, the scale invariant scattering framework should provide new elements also on this question. We will see that this is the case.

The $q$-state Potts model \cite{Wu} is defined on the lattice by the Hamiltonian
\EQ
{\cal H}=-\sum_{\langle i,j\rangle}J_{ij}\delta_{s_i,s_j}\,,\hspace{1cm}
s_i=1,2,\ldots,q\,,
\EN
where $s_i$ is the variable at site $i$, the sum is taken over nearest neighboring sites, and $J_{ij}$ are bond couplings. The model is characterized by the symmetry $\mathbb{S}_q$ corresponding to permutations of the $q$ values (``colors") that $s_i$ can take. In two dimensions the pure ferromagnet ($J_{ij}=J>0$) has a phase transition that is of the second order up to $q=4$ and becomes of the first order for $q>4$ \cite{Baxter}. The random bond model is obtained when the couplings $J_{ij}$ become random variables drawn from a probability distribution $P(J_{ij})$. The average over disorder is taken on the free energy
\EQ
\overline{F}=\sum_{\{J_{ij}\}}P(J_{ij})F(J_{ij})\,,
\EN
and is theoretically dealt with through the replica method \cite{EA}. This exploits that fact that, since $F=-\ln Z$, with $Z=\sum_{\{s_i\}}e^{-{\cal H}/T}$ the partition function, the relation
\EQ
\overline{F}=-\overline{\ln Z}=-\lim_{n\to 0}\frac{\overline{Z^n}-1}{n}
\EN
maps the problem onto that of $n\to 0$ replicas coupled by the disorder average. 

It was a key observation of \cite{random} that the replica method can be implemented in a truly exact form for two-dimensional systems at second order criticality within the scale invariant scattering approach of \cite{paraf}. This is possible  because the critical system is described by a two-dimensional Euclidean field theory that is the continuation to imaginary time of a conformally invariant quantum field theory with one space and one time dimension. The latter possesses a description in terms of massless particles\footnote{See \cite{fpu} for an overview on fields, particles and criticality in two dimensions.}, and infinite-dimensional conformal symmetry \cite{DfMS} forces infinitely many conserved quantities for the scattering of such particles. Scattering processes are then completely elastic (initial and final state are kinematically identical). In addition, since the center of mass energy is the only relativistic invariant of two-particle scattering and is dimensionful, scale invariance at criticality forces the scattering amplitude to be energy-independent. This in turn leads to a particularly simple form \cite{paraf,fpu} of the unitarity and crossing equations \cite{ELOP} satisfied by the amplitudes. 

\begin{figure}
\begin{center}
\includegraphics[width=10cm]{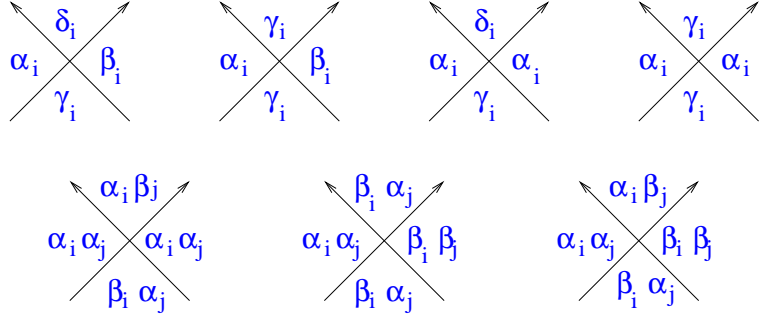}
\caption{Scattering processes in the replicated $q$-state Potts model. They correspond to the amplitudes $S_0$, $S_1$, $S_2$, $S_3$, $S_4$, $S_5$, $S_6$, in that order. Time runs upwards. Different latin indices indicate different replicas, and different greek letters for the same replica indicate different colors. 
}
\label{potts_ampl}
\end{center} 
\end{figure}

When applying this formalism to the Potts model, the key role is played by the color permutational symmetry $\mathbb{S}_q$. The symmetry is represented by particles $A_{\alpha\beta}$ ($\alpha,\beta=1,2,\ldots,q$; $\alpha\neq\beta$), which in the broken phase of the pure ferromagnet correspond to the kinks interpolating between degenerate ground states \cite{CZ}. More generally, the trajectories of the particles $A_{\alpha\beta}$ separate two regions characterized by different colors $\alpha$ and $\beta$; these excitations carry the basic representation of permutational symmetry that holds also at criticality \cite{paraf}, and for antiferromagnets \cite{DT}. The random case is obtained considering the replicated theory with excitations $A_{\alpha_i\beta_i}$, with $i=1,2,\ldots,n$ labeling the replicas \cite{random}. The trajectory of $A_{\alpha_i\beta_i}$ separates a region characterized by the colors $\alpha_1,\ldots,\alpha_n$ for the replicas $1,\dots,n$, respectively, from a region where replica $i$ has changed its color to $\beta_i$; the colors of the other replicas remain unchanged. Then the scattering processes allowed by the requirement of invariance under permutations of the replicas and permutations of the colors within each replica are those depicted in figure~\ref{potts_ampl}, with scattering amplitudes $S_0,S_1,\ldots,S_6$; only the replicas whose color changes in the scattering process are explicitly indicated in the figure. The first four amplitudes involve color change within a single replica, while the last three amplitudes introduce interaction among the replicas and are characteristic of the disordered case. For example, amplitude $S_4$ corresponds to an initial state with particles $A_{\alpha_i,\beta_i}$ and $A_{\beta_i,\alpha_i}$, both in replica $i$, and a final state with particles $A_{\alpha_j,\beta_j}$ and $A_{\beta_j,\alpha_j}$, both in replica $j$.

Crossing symmetry \cite{ELOP} relates amplitudes under exchange of time and space directions (by complex conjugation in the present context \cite{fpu}), and takes the form
\bea
S_0 = S_0^* &\equiv & \rho_0\,,
\label{c1}\\
S_1 = S_2^*&\equiv &\rho\,e^{i\varphi}\,,
\label{c2}\\
S_3 = S_3^*&\equiv &\rho_3\,,
\label{c3}\\
S_4 = S_5^*&\equiv &\rho_4\,e^{i\theta}\,,
\label{c4}\\
S_6 = S_6^*&\equiv &\rho_6\,,
\label{c5}
\eea
where we introduced 
\EQ
\rho_0, \rho_3,\rho_6, \varphi,\theta\in\mathbb{R}\,,\hspace{1cm}\rho, \rho_4\geq 0\,.
\label{constraints}
\EN
Unitarity of the scattering matrix encodes conservation of probability, and in the present case leads to the equations~\cite{random,DT2} 
\bea
&& \rho_3^2+(q-2)\rho^2+(n-1)(q-1)\rho_4^2 = 1\,,
\label{u1}\\
&& 2\rho\rho_3\cos\varphi+(q-3)\rho^2+(n-1)(q-1)\rho_4^2 = 0\,,
\label{u2}\\
&& 2\rho_3\rho_4\cos\theta+2(q-2)\rho\rho_4\cos(\varphi+\theta)+(n-2)(q-1)\rho_4^2 = 0\,
\label{u3}\\
&& \rho^2+(q-3)\rho_0^2 = 1\,,
\label{u4a}\\
&& 2\rho_0\rho\cos\varphi+(q-4)\rho_0^2 = 0\,,
\label{u4b}\\
&& \rho_4^2+\rho_6^2=1 \,,
\label{u5}\\
&& \rho_4\rho_6\cos\theta=0 \,.
\label{u6}
\eea
Notice that $q$ and $n$ enter the equations as parameters that can take real values. The continuation to real values of $q$ of the lattice model is known from the cluster expansion \cite{FK}.

When $n=1$ and the equations that still involve $\rho_4$ or $\rho_6$ are ignored, the equations (\ref{u1})--(\ref{u6}) reduce to those of the pure model \cite{paraf,DT}, as expected. It is also important to observe that $\rho_4=0$ yields $n$ non-interacting replicas, since it implies $S_4=S_5=0$ and, due to (\ref{u5}), $S_6=\pm 1$; in one spatial dimension scattering involves position exchange on the line, so that a scattering amplitude equal to $-$1 (resp. $1$) corresponds to non-interacting fermions (resp. bosons). It follows that $\rho_4=0$ corresponds to absence of disorder. 

The solutions of the unitarity equations (\ref{u1})--({\ref{u6}) have been listed in \cite{DT2}. They correspond to renormalization group fixed points with permutational symmetry of the $q$ colors and of the $n$ replicas. Here we directly consider the case $n=0$ that is relevant for quenched disorder. The solutions turn out to fall into three classes characterized by the values of $\rho_4$ as a function of $q$: solutions with $\rho_4=0$ correspond to the pure case, solutions with $\rho_4=1$ are always strongly disordered, solutions with $\rho_4$ depending on $q$ exhibit vanishing disorder at specific values of $q$. 

The equations for the pure case ($\rho_4=0$) are first of all characterized by the existence of a value $q_{\textrm{max}}=(7+\sqrt{17})/2=5.5615..$ above which no solution, i.e. no second order transition, exists \cite{DT}. It is also interesting that this value of $q_{\textrm{max}}$ is larger than the value 4 usually assumed from available lattice solutions \cite{Baxter,Baxter_square_AF}, and makes possible a second order transition in a $q=5$ antiferromagnet, a possibility for which lattice candidates have been considered in the literature on numerical grounds (see \cite{Deng,Huang}). 

Remaining to the pure case ($\rho_4=0$) of our interest, namely that of a ferromagnet, the equations yield a solution defined in the required interval $q\in[0,4]$. It reads \cite{paraf,DT}
\EQ
\rho_0=-1\,,\hspace{1cm}\rho=\sqrt{4-q}\,,\hspace{1cm}2\cos\varphi=-\sqrt{4-q}\,,\hspace{1cm}\rho_3=q-3\,,
\label{pure}
\EN
where we used the fact that the pure Ising model ($q=2$) is a theory of free fermions to fix the sign of $S_3=\rho_3$; this is the only physical amplitude at $q=2$, since the other amplitudes in the first row of figure~\ref{potts_ampl} involve more than two colors. 

It is known from the Harris criterion \cite{Harris} that weak disorder is relevant in the renormalization group sense when the critical exponent $\alpha=(2-2X_\varepsilon)/(2-X_\varepsilon)$ of the pure system is positive, i.e. when the energy density scaling dimension $X_\varepsilon$ of the pure system is smaller than 1. For the Potts model this condition is known to hold for $q\in(2,4]$ \cite{Nienhuis}, the upper extreme of the interval being the endpoint of second order criticality. For $q=2$, where $X_\varepsilon$ takes the value 1 characteristic of free fermions, weak disorder is marginally irrelevant \cite{DD} and does not produce a random fixed point with new critical exponents. This means that the random fixed point present for $q\in(2,4]$ can be studied perturbatively for $q\to 2^+$; the perturbative analysis was carried out in \cite{Ludwig,DPP}. On the other hand, the result of \cite{AW,HB} on the softening of first order transitions by disorder suggests that the presence of the random fixed point does not stop at $q=4$ but persists until $q=\infty$. This fixed point is expected to exist for real values of $q$, and was indeed studied numerically for $q$ noninteger in \cite{CJ}. In order to meet these requirements, the equations (\ref{u1})-(\ref{u6}) should possess a solution defined for all real values of $q\geq 2$, with $\rho_4=0$ and $\rho_3=-1$ at $q=2$. Such a solution exists and reads \cite{random}
\EQ
\rho _0=\cos\theta=0,\hspace{1cm} \rho =1, \hspace{1cm}  \rho _3=2 \cos\varphi =-\frac{2}{q}, \hspace{1cm}  \rho _4=\frac{q-2}{q} \sqrt{\frac{q+1}{q-1}}\,.
\label{ir}
\EN
Hence, scale invariant scattering provides the first analytic verification of the expectation coming from the combination of rigorous, perturbative and numerical results. 

\begin{figure}
\begin{center}
\includegraphics[width=10cm]{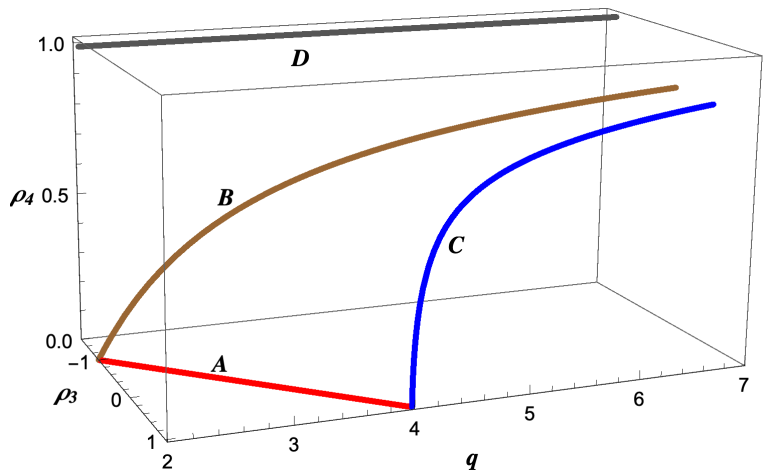}
\caption{Lines of renormalization group fixed points with $\mathbb{S}_q$ permutational symmetry associated to solutions of the equations (\ref{u1})-(\ref{u6}) with $n=0$. The lines $A$, $B$, $C$ and $D$ correspond to the solutions (\ref{pure}), (\ref{ir}), (\ref{uv}) and (\ref{nishimori})-(\ref{perc}), respectively. Absence of disorder corresponds to $\rho_4=0$, and the line of fixed points $A$ is that of the pure Potts ferromagnet. 
}
\label{soft_pd}
\end{center} 
\end{figure}

The picture, however, cannot yet be considered as complete. Indeed, the Harris criterion implies that for $q\in(2,4]$ there is a renormalization group flow from the pure ferromagnet (\ref{pure}) to the line of random fixed points (\ref{ir}). The latter, however, continues to be a line of infrared fixed points also for $q>4$, where the transition of the pure model is first order. If usual renormalization group mechanisms have to apply, the random model should possess for $q>4$ a line of unstable fixed points from which the flow towards solution (\ref{ir}) originates. In this case, the equations (\ref{u1})-(\ref{u6}) should admit a solution starting at $q=4$ and extending until $q=\infty$. As a matter of fact, such a solution exists and reads
\bea
&& \rho _0=-\frac{2}{|q^2-4 q+2|}\,, \hspace{1cm}
\rho=\frac{\sqrt{(q-4) (q^3-4 q^2+4 q-4)}}{|q^2-4 q+2|}\,,\nn
&& \rho _3=\frac{2(q-3)}{| q^2-4 q+2| }\,, \hspace{1cm}
\rho _4=\frac{(q-2)\sqrt{(q-4) (q-3) (q-1) q}}{(q-1)|q^2-4 q+2|}\,, \nn
&& 2 \cos\varphi=\frac{2 (q-4)}{\sqrt{(q-4) (q^3-4 q^2+4 q-4)}}\,,\hspace{1cm}\cos\theta=0\,.
\label{uv}
\eea
Notice that this solution coincides with that for the pure ferromagnet (\ref{pure}) at $q=4$ (see figure~\ref{soft_pd}). 

Both solutions (\ref{ir}) and (\ref{uv}) have $\cos\theta=0$ and for this reason possess a remarkable property observed in \cite{random}. The superposition  $\sum_{i,\gamma_i}A_{\alpha_i\gamma_i}A_{\gamma_i\alpha_i}$ belongs to the $\mathbb{S}_q$-invariant sector of the theory and scatters into itself with the phase
\EQ
S=S_3+(q-2)S_2+(n-1)(q-1)S_4\,.
\label{S}
\EN
It is then easy to check that for $\cos\theta=0$ (\ref{u3}) implies $\textrm{Im}\,S=0$ (i.e. $S=\pm 1$) at $n=0$. In other words, while the solutions (\ref{ir}) and (\ref{uv}) are generically $q$-dependent, their $\mathbb{S}_q$-invariant sector is $q$-independent. This property has no counterpart in absence of disorder and makes possible that the scaling dimensions of $\mathbb{S}_q$-invariant fields like the energy density $\varepsilon$ remain constant along the random critical line. In particular, it was observed in \cite{random} that this allows the exponent $\nu=1/(2-X_\varepsilon)$ to keep along the infrared stable critical line (\ref{ir}) the pure Ising value 1 that it takes at $q=2$, thus shedding light on the numerical puzzle we already mentioned. In the same way, $X_\varepsilon$ is expected to keep for $q>4$ along the unstable line (\ref{uv}) the value $1/2$ that it has at $q=4$ in the pure model. 

Notice that the expressions (\ref{ir}) and (\ref{uv}) formally coincide at $q=\infty$. However, it is easy to check that the amplitude (\ref{S}) calculated at $n=0$ takes for $q\geq 4$ the value $-1$ for (\ref{ir}) and the value $1$ for (\ref{uv}). This is possible because, due to the term $(q-2)S_2$, for both solutions this amplitude behaves for large $q$ as $q\cos\varphi$, with $\cos\varphi$ vanishing as $-1/q$ for (\ref{ir}) and as $1/q$ for (\ref{uv}). Hence, due to the peculiarity of the limit, the stable and unstable critical lines do not really merge as $q\to\infty$. 

The solution (\ref{uv}) is also defined for $q\in[2,3]$, with $\rho_4$ vanishing at the extrema of this interval. At $q=2$ it correponds to the Ising free fermion with $X_\varepsilon=1$; at $q=3$, however, it does not correspond to the Potts ferromagnet but, as shown in \cite{DT2}, to a theory of two free neutral fermions, again with $X_\varepsilon=1$. This difference at $q=3$ means that this branch enters a sector of the multidimensional parameter space that is not related to the phase diagram we are considering, and for this reason it is not shown in figure~\ref{soft_pd}.

Putting all together, the properties of solution (\ref{uv}) are consistent with the scenario that it provides the starting point for the renormalization group flow ending on the stable critical line (\ref{ir}) for $q>4$. This scenario suggests that for $q>4$ the second order transition may set in above a $q$-dependent disorder threshold. Below this threshold the transition would occur with finite correlation length but, to comply with the rigorous result of  \cite{AW}, without discontinuity in the energy density.

The stable critical line (\ref{ir}) is also expected to be the large distance limit of the flow originating from fixed points with stronger disorder. For the disorder distribution 
\EQ
P(J_{ij})=p\,\delta(J_{ij}-J)+(1-p)\,\delta(J_{ij}+J)\,,\hspace{.5cm}J>0\,,
\label{glass}
\EN
in which bonds are ferromagnetic with probability $p$ and antiferromagnetic with probability $1-p$, this strong disorder fixed point can be referred to as Nishimori-like fixed point, as a generalization of the Nishimori fixed point whose presence for $q=2$ can be argued from a lattice gauge symmetry \cite{Nishimori}. For $q=3$ this fixed point was studied numerically in \cite{SGH,JP2}. Within the space of solutions of equations (\ref{u1})-(\ref{u6}), the fixed points that never become weakly disordered as $q$ varies are in the class with $\rho_4=1$. This class contains two solutions defined for any $q$, one of which is completely $q$-independent and will be discussed in a moment. The other solution reads
\EQ
\rho _0=0,\hspace{1cm}\rho=\rho_4=1,\hspace{1cm} \rho_3=2\cos\varphi =-\sqrt{2},\hspace{1cm} 
2\cos\theta= -\frac{\sqrt{2}(q^2-2)}{\left(q^2-2 q+2\right)}\,, 
\label{nishimori}
\EN
and should correspond to the Nishimori-like fixed points. 

The completely $q$-independent strongly disordered solution reads
\EQ
\rho _0=0,\hspace{1cm}\rho=\rho_4=1,\hspace{1cm} \rho_3=2\cos\varphi=2\cos\theta =-\sqrt{2}\,,
\label{perc}
\EN
and its presence is expected. Indeed, for a dilute ferromagnet with disorder distribution
\EQ
P(J_{ij})=p\,\delta(J_{ij}-J)+(1-p)\,\delta(J_{ij})\,,\hspace{.5cm}J>0\,,
\label{dilute}
\EN
only clusters of spins connected by ferromagnetic bonds $J_{ij}=J$ contribute to the energy. At zero temperature this yields uniformly and independently colored spin clusters, and a total magnetization that vanishes unless there is an infinite cluster. Hence, there is for any $q$ a transition in the universality class of random percolation that is accounted for by (\ref{perc}). The zero-temperature fixed point can be shown to be unstable (see e.g. \cite{Cardy_book}), and originates a flow towards the fixed point (\ref{ir}). The solutions (\ref{nishimori}) and (\ref{perc}) differ only in the value of the parameter $\theta$ and are not distinguished in the parameter subspace of figure~\ref{soft_pd}. 

In summary, we have investigated the effect of quenched bond disorder on the first order transition of the two-dimensional $q$-state Potts model, for which the elimination of the latent heat has been known since the work of \cite{AW,HB}. We showed that the space of solutions of the exact fixed point equations implied by scale invariant scattering at criticality contains the expected stable critical line extending to arbitrary large values of $q$, but also a line of unstable fixed points from which the second order phase transition surface can originate for $q>4$, where the transition of the pure model is first order. These critical lines have been exactly located within the space of universal scattering parameters together with Nishimori-like and zero-temperature critical lines.

\vspace{1cm} 
\noindent 
\textbf{Acknowledgments.} We thank M.~Aizenman for an interesting discussion on ref.~\cite{AW}.


\vspace{1cm}
\begin{thebibliography}{99}
\bibitem{AW} M. Aizenman and J. Wehr, Phys. Rev. Lett. 62 (1989) 2503.
\bibitem{HB} K. Hui and A.N. Berker, Phys. Rev. Lett. 62 (1989) 2507. 
\bibitem{Ludwig} A.W.W. Ludwig, Nucl. Phys. B 330 (1990) 639.
\bibitem{DPP} V. Dotsenko, M. Picco and P. Pujol, Nucl. Phys. B 455 (1995) 701. 
\bibitem{CFL} S. Chen, A.M. Ferrenberg and D.P. Landau, Phys. Rev. Lett. 69 (1992) 1213; Phys. Rev. E 52 (1995) 1377. 
\bibitem{DW} E. Domany and S. Wiseman, Phys. Rev. E 51 (1995) 3074.
\bibitem{KSSD} M. Kardar, A.L. Stella, G. Sartoni and B. Derrida, Phys. Rev. E 52 (1995) R1269.
\bibitem{CJ} J. Cardy and J.L. Jacobsen, Phys. Rev. Lett. 79 (1997) 4063.

J.L. Jacobsen and J. Cardy, Nucl. Phys. B 515 (1998) 701.
\bibitem{random} G.~Delfino, Phys. Rev. Lett. 118 (2017) 250601.
\bibitem{paraf} G. Delfino, Annals of Physics 333 (2013) 1.
\bibitem{DL1} G. Delfino and N. Lamsen, JHEP 04 (2018) 077.  
\bibitem{DL2} G. Delfino and N. Lamsen, J. Stat. Mech. (2019) 024001.
\bibitem{DT2} G. Delfino and E. Tartaglia, J. Stat. Mech. (2017) 123303.
\bibitem{CB2} C. Chatelain and B. Berche, Phys. Rev. E 60 (1999) 3853.
\bibitem{OY} T. Olson and A. P. Young, Phys. Rev. B 60 (1999) 3428.
\bibitem{JP} J.L. Jacobsen and M. Picco, Phys. Rev. E 61 (2000) R13.
\bibitem{Jacobsen_multiscaling} J.L. Jacobsen, Phys. Rev. E 61 (2000) R6060(R).
\bibitem{AdAI} J.-Ch. Angl\`es d’Auriac and F. Igloi, Phys. Rev. Lett. 90 (2003) 190601.
\bibitem{Wu} F.Y. Wu, Rev. Mod. Phys. 54 (1982) 235.
\bibitem{Baxter} R.J. Baxter, Exactly Solved Models of Statistical Mechanics, Academic Press, London, 1982.
\bibitem{EA} S.F. Edwards and P.W.  Anderson, J. Phys. F 5 (1975) 965.
\bibitem{fpu} G. Delfino, Annals of Physics 360 (2015) 477.
\bibitem{DfMS} P. Di Francesco, P. Mathieu and D. Senechal, Conformal field theory, Springer-Verlag, New York, 1997.
\bibitem{ELOP} R.J. Eden, P.V. Landshoff, D.I. Olive and J.C. Polkinghorne, The analytic S-matrix, Cambridge, 1966.
\bibitem{CZ} L. Chim and A.B. Zamolodchikov, Int. J. Mod. Phys. A 7 (1992) 5317.
\bibitem{DT} G. Delfino and E. Tartaglia, Phys. Rev. E 96 (2017) 042137.
\bibitem{FK} P.W. Kasteleyn and E.M. Fortuin, J. Phys. Soc. Japan Suppl. (1969) 2611; Physica 57 (1972) 536.
\bibitem{Baxter_square_AF} R.J. Baxter, Proc. Roy. Soc. London A 383 (1982) 43.
\bibitem{Deng} Y. Deng, Y. Huang, J.L. Jacobsen, J. Salas, and A.D. Sokal, Phys. Rev. Lett. 107 (2011) 150601.
\bibitem{Huang} Y. Huang, K. Chen, Y. Deng, J.L. Jacobsen, R. Koteck ́, J. Salas, A.D. Sokal and J.M. Swart, Phys. Rev. E 87 (2013) 012136.
\bibitem{Harris} A.B. Harris, J. Phys. C 7 (1974) 1671.
\bibitem{Nienhuis} B. Nienhuis, J. Stat. Phys. 34 (1984) 731.
\bibitem{DD} V.S. Dotsenko and Vl. S. Dotsenko, Sov. Phys. JETP Lett. 33 (1981) 37; Adv. Phys. 32 (1983) 129.
\bibitem{Nishimori} H. Nishimori, Prog. Theor. Phys. 66 (1981) 1169.
\bibitem{SGH} E.S. Sorensen, M.J.P. Gingras and D.A. Huse, Europhys. Lett. 44 (1998) 504.
\bibitem{JP2} J.L. Jacobsen and M. Picco, Phys. Rev. E 65 (2002) 026113.
\bibitem{Cardy_book} J. Cardy, Scaling and renormalization in statistical physics, Cambridge, 1996.


\end{thebibliography}
\end{document}